\documentclass[aps,prl,floatfix,twocolumn,footinbib,amsmath,amssymb]{revtex4-1}
\usepackage{amsthm,color,latexsym,graphicx,array,multirow}

\newcommand{\eq}[1]{Eq.~\eqref{eq:#1}}

\newcommand{\ie}{{i.e.}}
\newcommand{\eg}{{e.g.}}

\newcommand{\beq}{\begin{eqnarray}}
\newcommand{\eeq}{\end{eqnarray}}

\newcommand{\gev}{~\text{GeV}}

\newcommand{\avg}[1]{\langle#1\rangle} 

\newcommand{\sla}[1]{
   \setbox0=\hbox{$#1$} \dimen0=\wd0
   \setbox1=\hbox{/} \dimen1=\wd1
   \ifdim\dimen0>\dimen1
      \rlap{\hbox to \dimen0{\hfil/\hfil}}
      #1
   \else
      \rlap{\hbox to \dimen1{\hfil$#1$\hfil}}
      /
   \fi}

\newcommand{\dmin}{d^{\text{min}}}

\newcommand{\kt}{\text{k}_{\text{T}}}
\newcommand{\CA}{\text{C/A}}
\newcommand{\Ntree}{N_{\text{tree}}}

\newcommand\Ref\cite

\newcommand{\pair}[2]{\langle #1  #2 \rangle}

\newcommand{\Qjet}{\text{Q-jet}}
\newcommand{\Qjets}{\text{Q-jets}}

\newcommand{\Qpruning}{Qjet-{pruning}}

\begin{document}

\title{\Qjets: A Non-Deterministic Approach to Tree-Based Jet Substructure}

\author{Stephen D. Ellis}
\email{sdellis@u.washington.edu}
\affiliation{Department of Physics, University of Washington, Seattle WA, 98195}
\author{Andrew Hornig}
\email{ahornig@u.washington.edu}
\affiliation{Department of Physics, University of Washington, Seattle WA, 98195}
\author{David Krohn}
\email{dkrohn@physics.harvard.edu}
\affiliation{Department of Physics, Harvard University, Cambridge MA, 02138}
\author{Tuhin S. Roy}
\email{tuhin@u.washington.edu}
\affiliation{Department of Physics, University of Washington, Seattle WA, 98195}
\author{Matthew D. Schwartz}
\email{schwartz@physics.harvard.edu}
\affiliation{Department of Physics, Harvard University, Cambridge MA, 02138}
\date{\today}

\begin{abstract}

Jet substructure is typically studied using clustering algorithms, such as $k_T$, which arrange the jets' constituents into trees.
Instead of considering a single tree per jet, we propose that multiple trees should be considered, weighted by an appropriate metric. Then each jet in each event produces a distribution for an observable, rather than a single value.
Advantages of this approach
include: 1) observables have significantly increased statistical stability; and, 2) new observables, such as the variance of the distribution, provide new handles for signal and background discrimination.
For example, we find that employing
a set of trees
substantially reduces the observed fluctuations in the pruned mass distribution,
enhancing the likelihood
of new particle discovery for a given integrated luminosity. Furthermore, the
resulting pruned mass distributions for (background) QCD jets are found to be substantially wider
than that for (signal) jets with intrinsic mass scales, {\eg} boosted $W$ jets.  A cut on this width yields a substantial enhancement in
significance relative to a cut on the standard pruned jet mass alone. In particular the luminosity needed for a given significance requirement decreases by a factor of two relative to standard pruning.
\end{abstract}
\maketitle

To develop intuition about high-energy collisions like those at the LHC  it is often helpful to think of an event as being produced by a multi-stage process. In this picture, a short distance scattering produces a few hard partons.
The partons then shower soft and collinear QCD radiation.
Finally, at long distances, the (colored) partons bind into the (color singlet) hadrons that we observe in the detector.
This parton-shower picture explains how clusters of nearby final-state particles, called jets, defined by a jet algorithm,
can reveal something about the short-distance physics. Simulations of the parton shower produce
events which, with sufficient tuning, exhibit remarkable agreement with collider data for nearly any conceivable infrared safe observable.

If one takes the parton-shower picture literally, the constituents of a jet 
arise from  a shower-like series of  $1\to 2$ splittings producing a ``tree'' structure.  Since the shower model for QCD is dominated by
soft and collinear splittings, any deviation from this behavior could indicate the presence of contamination within the jet, or might
indicate that the jet is not purely of QCD origin (\eg, it could come from a boosted heavy particle).
Thus, by associating trees (by ``trees,'' we mean ``clustering histories'') to jets one can obtain useful information, and  indeed this is the basis for much of the  work in the field of jet substructure (see Ref.~\cite{Abdesselam:2010pt,*Almeida:2011ud,*Salam:2009jx,*2012arXiv1201.0008A} for a review).

The association of a tree to a jet naturally emerges from the parton-shower picture.
In the parton shower, soft and collinear radiation
is emitted in a particular sequence: a $p_{\rm T}$-ordered shower
builds a tree by adding on emissions in decreasing order of transverse momentum, while an angular ordered  shower
adds emissions in a sequence of decreasing angle.  The recombination jet algorithms try to match this behavior.
The $\kt$ algorithm~\cite{Ellis:1993tq,*Catani:1993hr} assembles a jet in increasing order of the (relative) $\kt$ metric that depends
on both angle and the magnitude of the momentum, and the Cambridge/Aachen ($\CA$)~algorithm~\cite{Wobisch:1998wt,*Dokshitzer:1997in}
assembles in increasing order of angle.  Both can be viewed as a reasonable guess for the showering sequence history.

One problem with thinking of jet algorithms as reversing the parton shower is that the parton shower
is not invertible -- a given set of four-momenta of final state particles could have evolved through a multitude of intermediate trees.
In this paper we propose a way to account for the non-invertible nature of the parton shower by associating to each jet a set of trees
instead of a single tree.

Related ideas have been discussed in the past.  
Long ago a probabilistic approach was used to improve the
behavior of seeded jet algorithms~\cite{Giele:1997ac}.
More recently, it has been shown that combining even highly correlated observables,
such as jet masses arising from different grooming techniques~\cite{Gallicchio:2010dq,*Cui:2010km,*Soper:2010xk}, can improve
discovery significance.  In addition, Ref.~\cite{Soper:2011cr} considered associating multiple
trees to a jet to compare with models of  showering in signal and
background processes, and Ref.~\cite{Volobouev:2011zz} proposed a
measure of jet {\it fuzziness} to gauge the ambiguity in jet
reconstruction.
However, our approach is fundamentally different from these previous studies.
We are interested in observables constructed from a  \textit{distribution} of trees for each jet
in each event.
For instance, we will show that by averaging tree-based observables over the trees for each jet,
their statistical stability can be substantially improved.

Associating a set of trees to a jet  would not
be feasible if one had to consider {\it every} tree which could be formed from a given set of final state four-momenta in a jet.
Fortunately, good approximations to such distributions obtained using every tree can be captured through a procedure analogous to Monte-Carlo integration, allowing us to use a very small fraction of the trees.
This is possible beause infrared and collinear safe jet observables must be insensitive to small reshufflings of the momenta, implying that large classes of trees give very similar information.

The algorithm we propose assembles a tree via a series of $2\rightarrow 1$ mergings:
\begin{enumerate}
\item At every stage of clustering, a set of weights $\omega_{ij}$ for all pairs $\pair{i}{j}$ of the four-vectors is computed, and a probability $\Omega_{ij} = \omega_{ij}/N$, where $N = \sum_{\pair{i}{j}} \omega_{ij}$, is assigned to each  pair.
\item A random number is generated and used to choose a pair $\pair{i}{j}$ with probability  $\Omega_{ij}$. The chosen pair is merged, and the procedure is repeated until all particles all clustered.
\end{enumerate}
This algorithm directly produces trees distributed according to their weight $\prod_{{\rm mergings}}\Omega_{ij}$. To produce a distribution of trees for each jet, this algorithm is simply repeated $\Ntree$ times (not necessarily yielding $\Ntree$ \textit{distinct} trees).
Note that any algorithm which modifies a tree during its
construction (\eg, jet pruning) can be adapted to work with this procedure, as demonstrated below.

One particularly interesting class of weights is given by
\begin{equation}
\label{eq:shw}
\omega_{ij}^{(\alpha)} \equiv \exp \left\{ - \alpha  \frac{(d_{ij} - \dmin)}{\dmin}  \right\}.
\end{equation}
with $\alpha$ a real number we call {\it rigidity}.
Here, $d_{ij}$ is the jet distance measure for the $\pair{i}{j}$ pair, \eg,
\begin{equation}
\label{eq:dij_defns}
d_{ij} =
\begin{cases}
&d_{\kt} \equiv {\rm min}\{ p_{Ti}^2, p_{Tj}^2 \}\Delta R_{ij}^2 \\
&d_{\CA} \equiv \Delta R_{ij}^2
\end{cases}
\,,\end{equation}
where $\Delta R_{ij}^2=\Delta y_{ij}^2  + \Delta \phi_{ij}^2$,
and $\dmin$ is the minimum over all pairs at this stage in the clustering.
Note that  with this metric, our algorithm reduces to a traditional clustering algorithm
when $\alpha \rightarrow \infty$, \ie, in that limit the \textit{minimal} $d_{ij}$ is always chosen.  In this sense, it is helpful to think of the traditional, single tree algorithm as the ``classical'' approach, with $\alpha \sim 1/\hbar$ controlling the deviation from the ``classical'' clustering behavior.
With this analogy, we call the trees constructed in this non-deterministic fashion $\Qjets$ (``quantum'' jets).

\begin{figure}
\includegraphics[width=0.45\textwidth]{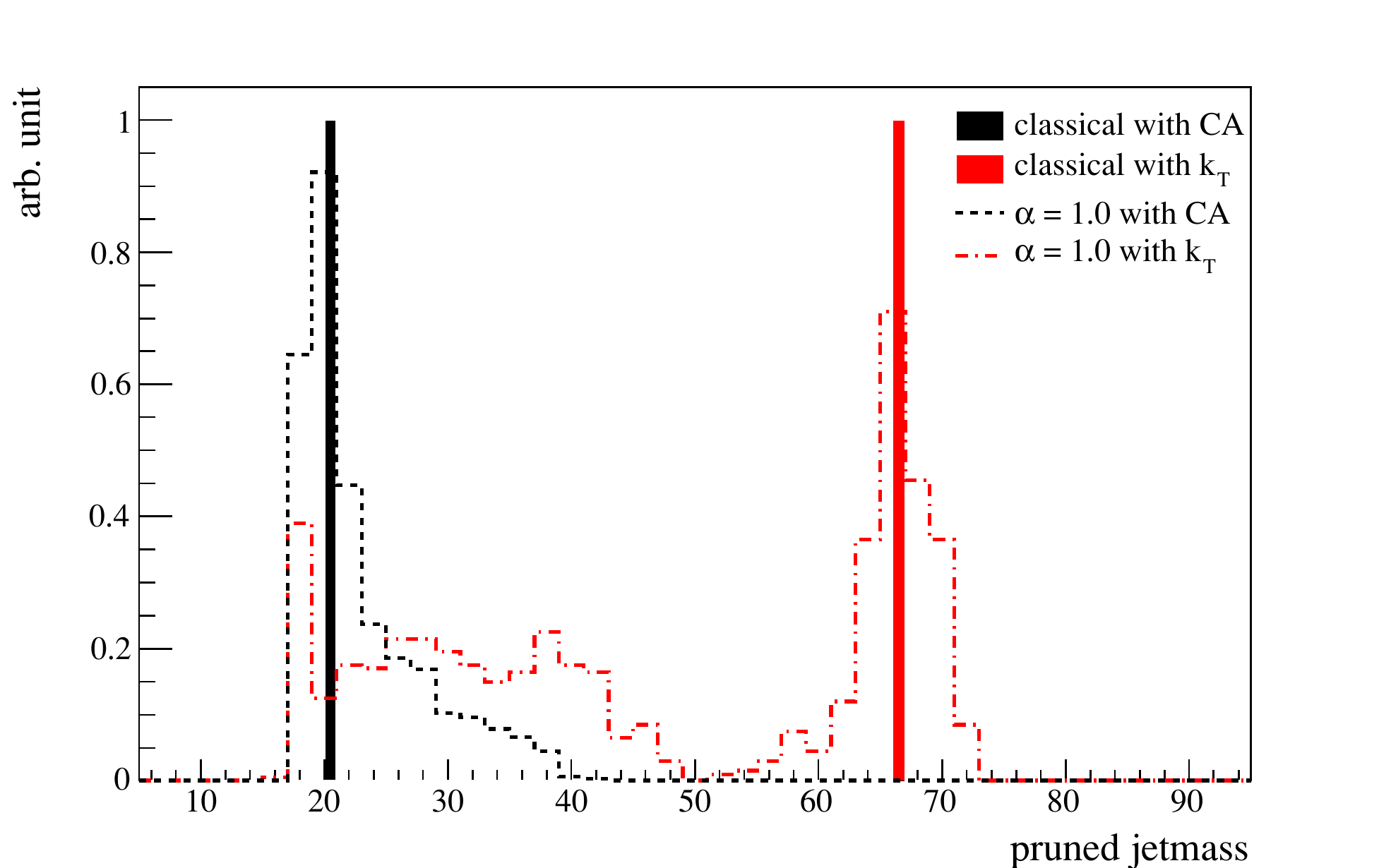}
\includegraphics[width=0.45\textwidth]{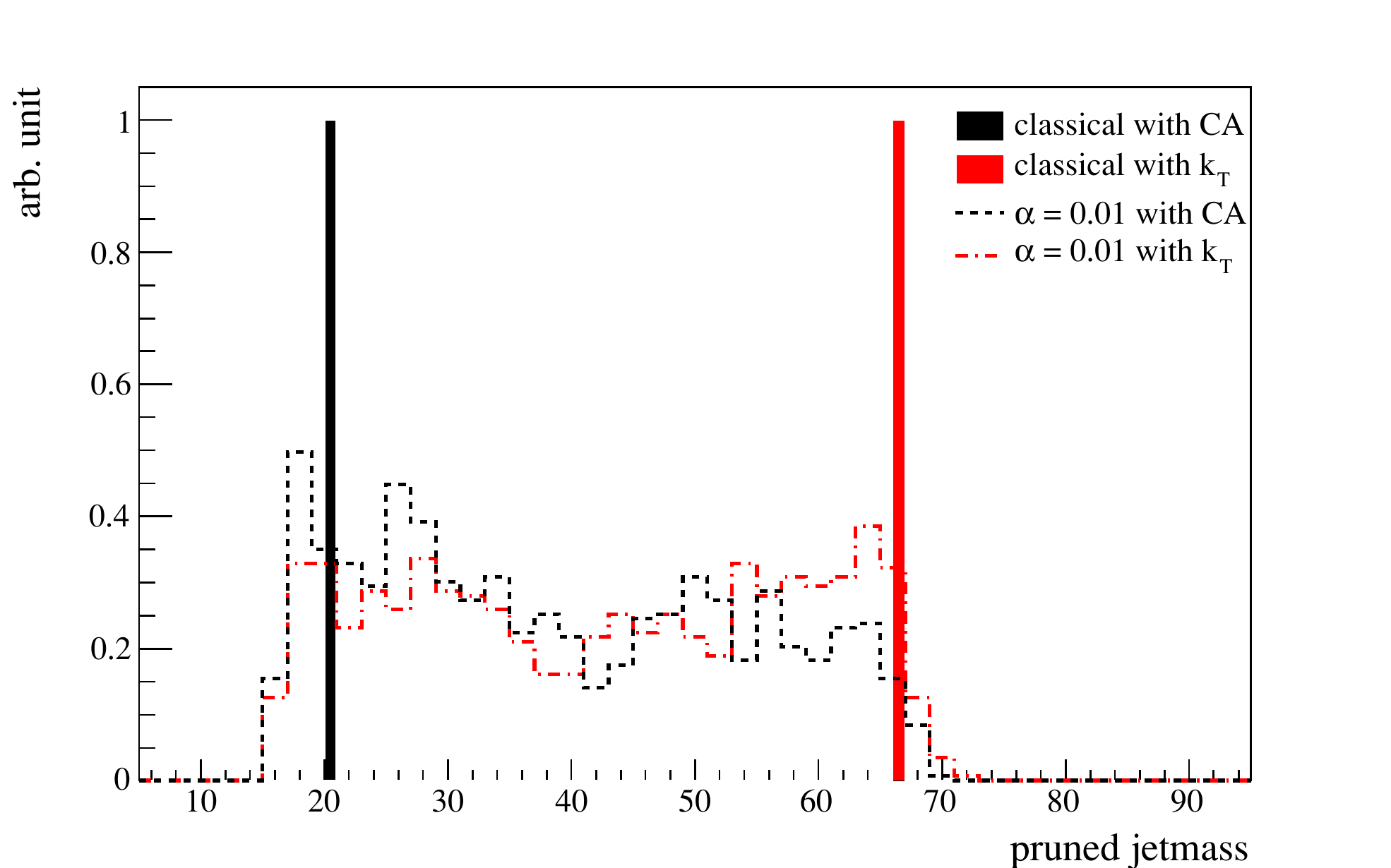}
\caption{\label{fig:rigidity} Distribution of pruned jet mass for a single
QCD-jet with $p_T \sim 500\gev$.  The black and red solid lines show the classical pruned masses when $\CA$ and $k_T$ 
algorithms are used to cluster the jet. The black and dashed (red and dot-dashed) line shows the pruned jet mass distribution of 1000 trees (constructed
from the same jet in the same event), when the $\CA$ ($\kt$) measure is used in Eq.~\eqref{eq:shw}. 
These distributions result from clusterings with rigidity $\alpha=1.0$ (top) and $\alpha=0.01$ (bottom).}
\vspace{-2mm}
\end{figure}

In order to get the most information out of the $\Qjets$, it is logical
to consider observables which are sensitive to the ordering of the clusterings in the
tree. One such observable is the pruned jet mass, which we will use as our illustrative example.
As described in Ref.~\cite{Ellis:2009me,*Ellis:2009su} pruning is one of the jet grooming tools~\cite{Butterworth:2008iy,*Kaplan:2008ie,*Krohn:2009th}. 
It is used  to sharpen signal and reduce background when considering boosted heavy objects.
The basic idea is to move along the tree and try to discard radiation which is soft and not collinear,
and therefore likely to represent contamination from a part of the event in which
we are not particularly interested (like the underlying event).
In detail, if a step in the clustering would merge particles $i$ and $j$ which satisfy
\begin{equation}
  \label{eq:4}
  \begin{split}
    & z_{ij} \equiv  \frac{ \text{min} \left( p_{T_i}, p_{T_j} \right) }{| \vec{p_{T_i}}+ \vec{p_{T_j}}| }  \ < \ z_\text{cut}  \qquad\text{and}\\
    & \Delta R_{ij} > D_\text{cut}  \, ,
  \end{split}
 \end{equation}
then the merging is vetoed and the softer of the two four-momenta is discarded.
In the specific analysis described here we take $z_\text{cut} = 0.1$ and $D_\text{cut} = m_\text{jet}/p_\text{jet}$,
which are typical cuts for the $\CA$ algorithm.

We apply this pruned $\Qjets$ procedure to samples of simulated boosted $W$ (signal) and QCD (background) jets generated with
\texttt{Pythia v6.422}~\cite{Sjostrand:2006za} with $p_T$-ordered showers using
the Perugia 2011 tunes~\cite{Skands:2010ak} and assuming a $7~{\rm TeV}$ LHC.
In lieu of detector simulation we group the visible output of Pythia into massless $\Delta \eta \times \Delta \phi = 0.1 \times 0.1$
``calorimeter cells'' (with $\vert\eta\vert < 5$), preserving the energy and the direction to the cell.
The cells with energy bigger than $0.5\gev$ become the inputs to the initial jet-finding algorithm (small alterations to this cut have no appreciable impact on our results). To find the initial jets we use
the anti-$\kt$ algorithm~\cite{Cacciari:2008gp} with $R = 0.7$ as implemented in \texttt{Fastjet v2.4.2}~\cite{Cacciari:Fastjet,*Cacciari:2005hq,*Cacciari:2011ma} and require $p_T^\text{jet} \ge 500\gev$\@.
Once a jet is identified, the cells clustered in the jet become input to the \Qpruning ~algorithm.
A fastjet plugin with this implementation of $\Qjets$ is available at \url{http://jets.physics.harvard.edu/Qjets}.

Consider first a single QCD jet from the sample described above. 
Fig.~\ref{fig:rigidity}
exhibits the pruned mass distribution for this jet obtained with the classical procedure for both
$\kt$ and $\CA$ pruning (the 2 vertical lines) and with $\Ntree = 1000$ using both the $\kt$ and $\CA$ metrics for $d_{ij}$ in  \eq{shw}.
The curves illustrate the dependence on the form of $d_{ij}$, as well as on the value of the
rigidity parameter $\alpha$. 
The upper panel is for $\alpha = 1.0$ where the trees are confined to stay close to the classical tree and the pruned masses likewise stay near the corresponding 
classical result.  For small enough $\alpha$
(say, $\alpha \lesssim 0.1$), a broad spectrum of trees is sampled.
This is shown in the lower panel of Fig.~\ref{fig:rigidity} for $\alpha = 0.01$, where
the distributions generated with the $\kt$ and $\CA$~definitions of the distance $d_{ij}$
look similar, and have little correspondence with the classical results.
This suggests that for a small enough rigidity parameter pruned $\Qjets$ become independent of the choice of distance measure used; they are therefore more likely to be characterizing physical features of an event rather than artifacts of using a particular jet algorithm.
%

We will now discuss two fundamentally different ways in which the discovery potential (\eg for finding boosted $W$ jets on top of their QCD background) can be enhanced using $\Qjets$:
\begin{itemize}
\item Observables have smaller statistical variation.  Even for the same number of background jets,
the use of $\Qjets$ reduces the background fluctuations $\delta B$ and increases the discovery potential $S/\delta B$,
where $S$ and $B$ are the numbers of signal and background jets in
the signal window and $\delta B$ denotes the fluctuation in $B$.
\item Qualitatively new observables, which depend on there being a distribution of trees for each jet, can now be considered.
For example, we define below a powerful observable we call {\it volatility} which measures the width of the pruned $\Qjet$ mass distribution for each jet, something inaccessible to a classical jet algorithm
\end{itemize}

To quantify the first of these points, we consider a large number of pseudo-experiments, each of which analyses $N_J$ jets,
with $N_J$ taken from a Poisson distribution with mean
 $\avg{N_J}$.
With a classical jet algorithm we can extract a significance by counting, in each pseudo-experiment, the number $S$ and $B$, of $W$ jets or QCD jets respectively, with pruned mass in a signal window, say between $70 - 90\gev$. The significance is then given by $\avg{S} /\delta B$, where $\avg{S}$ is the average over the pseudo-experiments of the number of signal events in the window and  $\delta B$ is the RMS fluctuation of $B$ over those pseudo-experiments. 
As expected $\avg{S}$ and $\avg{B}$ are proportional to $\avg{N_J}$, while $\delta S$ and $\delta B$ vary with $\sqrt{\avg{N_J}}$.  In addition to looking at $\avg{S}/\delta B$, we can also look at  the RMS fluctuations in the average pruned \Qjet~mass of the signal jets, $\delta \avg{m}$, averaged over the signal jets in the signal window for each pseudo-experiment. This tells us the statistical uncertainty with which the $W$ mass could be measured from these events.

With $\Qjets$, we can do something more sophisticated. 
Instead of the contribution of a given jet to $S$ or $B$ being 1 or 0 depending on whether the pruned mass is in the signal window or not, the contribution of the jet is now a rational number between 0 and 1, given by the fraction of the $\Ntree$ pruned masses that fall in the signal mass window.
This is a way of reducing
the contribution from events which are less signal like, without discarding them completely.
 In the limit $\alpha \to \infty$,
this reduces to the classical measure, but for finite $\alpha$, we expect an improvement in both significance and in $\delta \avg{m}$.

\begingroup
\squeezetable
\begin{table}
\begin{tabular}{|c|c|ccccc|}
\hline
\multirow{2}{*}{}&Vol. & \multicolumn{5}{c|}{Rigidity}\\
& cut (${\cal V_\text{cut}}$)& $\alpha = 0$&$\alpha = 0.01$& $\alpha = 0.1$ &
$\alpha = 1$& $\alpha = 100$ \\
 \hline
 \hline
\multirow{5}{*}
{{\large $\frac{{\avg{S}/\delta B}|_\text{Q}}{{\avg{S}/\delta B}|_\text{cl}}$}}
&{\bf None} & 1.07(1) & 1.13(1)& 1.18(1) & 1.14(1)& 1.06(1)\\
 & 0.05 & 1.43(4) & 1.44(3)& 1.39(3) & 1.27(1) & 1.08(1)\\
 & 0.04 & 1.51(4) & 1.45(4)& 1.39(3) & 1.29(3) & 1.10(1)\\
 & 0.03 & 1.51(2) & 1.45(3)& 1.37(4) & 1.35(2) & 1.10(1)\\
&  0.02 & 1.28(5) & 1.24(3) & 1.28(3) & 1.36(3)&1.13(1)\\
 \hline
 \hline
\multirow{5}{*}
{ {\large $\frac{\delta \avg{m}|_\text{cl}}{\delta\avg{m}|_\text{Q}}$} }
& {\bf None} &  1.32(2)& 1.31(2) & 1.25(2) & 1.10(2) &1.03(1)\\
 & 0.05 & 0.80(1) & 0.80(1) & 0.81(1) & 0.96(1) & 1.01(1)\\
 & 0.04 & 0.62(3) & 0.69(3) & 0.71(2) & 0.93(1) & 1.00(1)\\
 & 0.03 & 0.56(4) & 0.57(5) & 0.60(4) & 0.87(1) & 0.98(1)\\
 &  0.02 &0.48(7) & 0.49(7) & 0.50(7) & 0.77(2)&0.95(1)\\
 \hline
\end{tabular}
 \caption{The improvement found in various measurements performed
using the \Qjet~procedure compared to the classical pruning result, for a range of values of the rigidity
parameter ($\alpha$) and subject to a set of volatility cuts ($\cal {V} \leq \cal {V}_\text{cut}$).
The first set of rows exhibit the discovery potential  $\avg{S}/ \delta B$, 
while the second shows the
average jet mass fluctuation $\delta \avg{m}$.
In both cases results greater than unity indicate improvement over the classical
pruning procedure (see the text for further discussion).
For all quantities, the approximate statistical uncertainty
for the last digit is shown in parenthesis. 
}
 \label{table:results}
 \vspace{-6mm}
\end{table}
\endgroup

For numerical analysis  we use the $\CA$ algorithm for both the classical and $\Qjets$ cases and take $\Ntree = 256$.
(We find that the results saturate for $\Ntree \gtrsim 50$).
We present results in Table \ref{table:results} as ratios of the $\Qjets$ result to the classical result, indicating the improvement in significance and mass uncertainty we can expect.
These ratios should be independent of $\avg{N_J}$ and so we determine statistical uncertainties 
by fitting to results for $\avg{N_J} =5,10,15$ and $20$. The approximate statistical uncertainties are shown in parenthesis and apply to the last digit.
We performed $10^4$ pseudo-experiments, expecting ${\cal O}(1\%)$ statistical fluctuations from this procedure.

The first set of rows in Table \ref{table:results} display measurements of the discovery potential $\avg{S}/ \delta B$ compared to the results with classical pruning. Focus on the rows
labeled ``none'' for now (volatility is explained below).
Since this quantity scales as $\sqrt{\cal L}$, the square of the number in the Table can be interpreted as an effective luminosity improvement due to employing the
\Qjet~procedure.  For example, for $\alpha =0.1$ the number $1.18$ means an effective increase in the luminosity by $(1.18)^2 - 1 = 39\%$.
 Larger $\alpha$ values confine the range of trees and yield results very near the classical pruning results, \ie, 
$\frac{\avg{S}}{\delta B}\big|_\text{Q} \to \frac{\avg{S}}{ \delta B} \big|_\text{cl}$. 
Smaller $\alpha$ values 
($\alpha < 0.1$, with a much broader range of trees) also tend to degrade (decrease) the discovery potential.

The second set of rows exhibit the average jet mass fluctuation $\frac{\delta \avg{m}|_\text{cl}}{\delta \avg{m}|_\text{Q}}$ (note classical
over $\Qjets$ here).  Values greater than unity mean that the mass can be measured more precisely with the \Qjet~ procedure for the same luminosity.
Note that  there is continuing improvement in $\delta\avg{m}$ as $\alpha$ decreases.
That we get sensible results for (\ie with a flat distance measure) is presumably because pruning is relatively insensitive to which tree we assign; even for physically unlikely clusterings, the hard radiation that reconstructs the mass is typically not pruned away.

\begin{figure}
\includegraphics[width=0.45\textwidth]{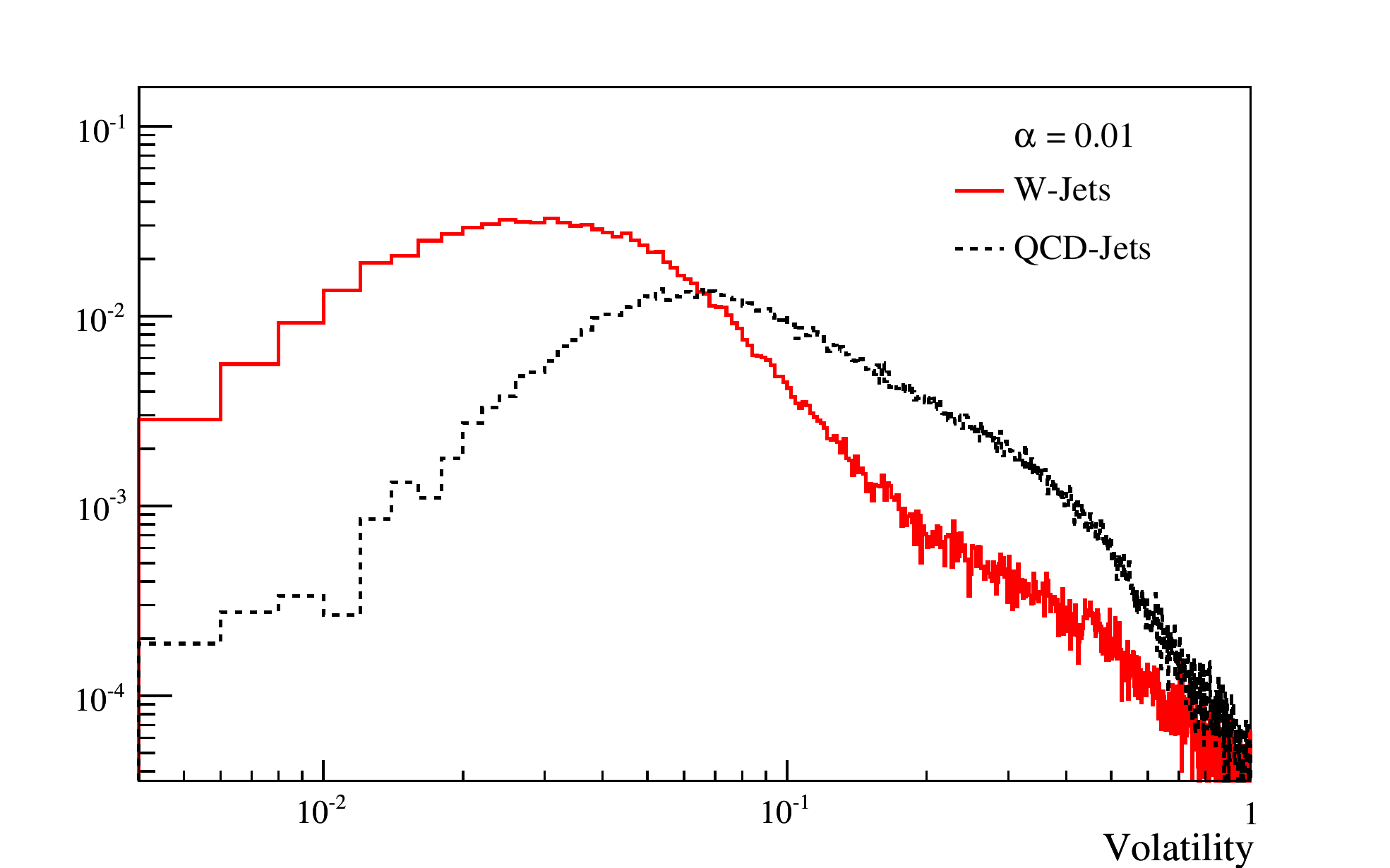}
\caption{Distribution of volatility for signal (boosted $W$-jets) and background (QCD jets) using a rigidity $\alpha=0.01$. 
\label{fig:volatility}}
\vspace{-4mm}
\end{figure}

The second way we have considered using $\Qjets$ is in constructing qualitatively
new types of observables.
As an example, consider the {\bf volatility} of a jet, defined by
\begin{equation}
  \label{eq:volatility}
  \mathcal{V} = \Gamma/\avg{m} \; ,
\end{equation}
where $\Gamma \equiv \sqrt{\langle m^2\rangle - \langle m\rangle^2}$ and $\avg{m}$ are the RMS deviation and the mean of the pruned jet mass distribution for a single jet.
The distribution of volatility for signal and background $\Qjets$ with $\alpha = 0.01$ is shown in Fig.~\ref{fig:volatility}. 
We see that $W$ jets have a lower volatility than QCD jets. This is easily
understood, since the $W$ jets have an intrinsic physical mass scale, while the QCD jets do not.
Cutting on volatility, $\cal {V} \leq \cal {V}_\text{cut}$ can therefore improve significance in
a boosted $W$ search. 
The  improvement is given in Table 1 for various values of  $\cal {V}_\text{cut}$.

\begin{figure}
\includegraphics[width=0.47\textwidth]{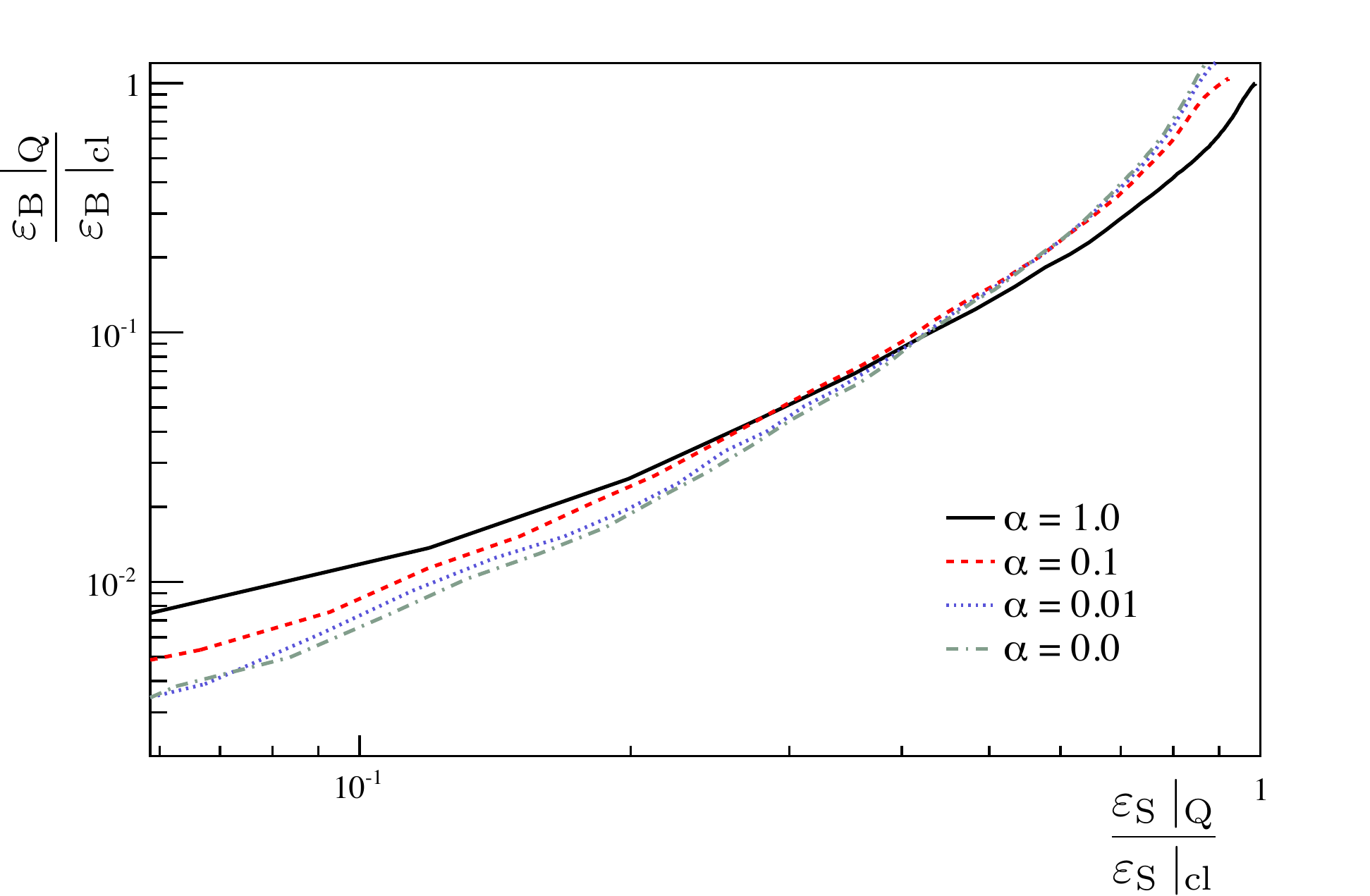}
\caption{The background versus signal efficiencies corresponding to a cut on volatility, for various $\alpha$'s,
as compared to the classical pruning result.
\label{fig:effs}}
\vspace{-4mm}
\end{figure}

The efficiencies for a volatility cut on signal and background are shown in Fig.~\ref{fig:effs}.
These efficiencies are defined as the fraction of  the $\Qjets$ that yield a pruned mass in the mass bin after the volatility cut. We plot them normalized to the classical results ($\alpha=\infty$ with
no volatility cut). In the limit $\alpha \to \infty$ the curve collapses to the point (1,1).
The upper right region of the plot corresponds to large values of $\cal {V}_\text{cut}$,
\ie, effectively no volatility cut. 
We find that the largest signal significance is obtained for a volatility cut of approximately $0.03$, where for $\alpha$ near zero we achieve a relative $\avg{S}/\avg{B}$ of $\sim 9$ and a relative $\avg{S}/\delta B$ improvement of $\sim 1.5$ (the square of this number is the factor of two quoted in the Abstract). This corresponds to the neighborhood of the point $(0.25, 0.03)$ in Fig.~\ref{fig:effs}.
Finally we note that
the precision of the mass measurement, shown in the lower rows in the table, is somewhat degraded
by placing a cut on the volatility.  This should not be a surprise as the cut discards some of the signal jets. A more comprehensive discussion of the statistics and of volatility will
be given in~\cite{Ellis:2012aa}.

In this paper, we have shown that it can be advantageous to consider a large number of trees constructed from the same jet in a single event,
rather than a single tree as is done in traditional clustering algorithms. 
Although this paper has focused on tree-based observables,
the $\Qjets$ idea, of using non-determinism in event analysis, 
 can naturally be applied in many other ways.
 Indeed, most observables, including 
jet substructure observables, such as jet masses, moments, pull~\cite{Gallicchio:2010sw}, jet shapes~\cite{Ellis:2010rwa, *Gallicchio:2011xq}, {\it etc.}, 
as well as more global observables, such as the number, distribution and 4-momenta for the jets in an event, work by trying to make the best guess at which properties of which final state particles tell us the most information about the underlying physics. 
The basic idea for $\Qjets$ is that there is an inherent ambiguity in this best guess, both due to there not being a precise correspondence between
final state particles and underlying physics, and due to our poor ability to extract that correspondence even if it were well-defined (as in a color
singlet decay, for example). Thus, it would be natural to consider multiple interpretations of any observable, to see whether getting away from
the best guess can give us more robust information about the underlying physics, as it has with the tree-based substructure considered here. 
In will be interesting to see in future work how far this non-deterministic approach can be pushed.

SDE, AH, and TSR were supported in part by US Department of Energy under contract number
DE-FGO2-96ER40956. MDS was supported in part by the Department of Energy, under 
grant DE-SC003916. 
DK was supported in part by a Simons postdoctoral fellowship and by an LHC-TI travel grant. AH, DK, and TSR were supported in part by the KITP, where a portion of this work was completed, under National Science Foundation under Grant No. PHY05-51164. Some computations were performed on the Odyssey cluster at Harvard University. 

 \vspace{-6mm}

\bibliography{references}
\bibliographystyle{apsrev4-1}

\end{document}